\documentclass[twocolumn]{webofc}
\usepackage[varg]{txfonts}   
\usepackage{siunitx}
\usepackage[font=small,labelfont=bf, justification=justified, format=plain]{caption}

\newcommand{\Hbar}{$\overline{\text{H}}$\xspace}
\newcommand{\pbar}{$\overline{\text{p}}$\xspace}
\newcommand{\pos}{$\text{e}^+$\xspace}

\begin{document}
\title{A Rydberg hydrogen beam for studies of stimulated deexcitation}
%
%
\author{
\firstname{Tim} \lastname{Wolz}\inst{1}\fnsep\thanks{\email{tim.wolz@cern.ch}} \and
\firstname{Maxime} \lastname{Allemand}\inst{2} \and
\firstname{Daniel} \lastname{Comparat}\inst{3} \and
\firstname{Jules} \lastname{Cras}\inst{3} \and
\firstname{Carina} \lastname{Killian}\inst{4} \and
\firstname{Chloé} \lastname{Malbrunot}\inst{1}\fnsep\thanks{Present address: TRIUMF, 4004 Wesbrook Mall, Vancouver, BC, V6T 2A3, Canada} \and
\firstname{Fredrik} \lastname{Parnefjord Gustafsson}\inst{4} \and
\firstname{Martin} \lastname{Simon}\inst{4} \and
\firstname{Christophe} \lastname{Siour}\inst{3} \and
\firstname{Eberhard} \lastname{Widmann}\inst{4} \, on behalf of the AEGIS and ASACUSA-CUSP collaboration
}

\institute{Physics Department, CERN \and
\'Ecole normale supérieure Paris-Saclay \and
Université Paris-Saclay, CNRS, Laboratoire Aimé Cotton, Paris \and
Stefan Meyer Institute, Vienna 
}

\abstract{
We present a Rydberg hydrogen beamline developed to commission techniques of stimulated deexcitation for application in antihydrogen experiments at CERN's Antiproton Decelerator. The stimulation of spontaneous decay is a key technology to enhance the number of ground-state anti-atoms available in a beam toward precision spectroscopy and gravity measurements.
}

\maketitle

\section{Introduction}
\label{intro}

Atomic antimatter systems are synthesized at CERN's Antiproton Decelerator facility to perform stringent tests of CPT symmetry and gravity. Antihydrogen (\Hbar) atoms can be currently produced relying on a resonant charge-exchange of laser excited positronium (a short-lived bound state of an electron and a positron (\pos)) and trapped antiprotons (\pbar) \cite{AEGIS_production}. Alternatively, experiments rely on a three-body-reaction in a \pbar-\pos plasma involving an antiproton and two positrons, one of which falls into a bound state with the antiproton and the other one carries away the atomic binding energy \cite{ALP_accumulation, ASACUSA_prod}. In both cases, \Hbar atoms are formed in a distribution of highly excited Rydberg quantum states exhibiting radiative lifetimes up to several milliseconds. In particular the numerous high angular momentum states are very long-lived while experiments require \Hbar atoms in their ground state.

Unless employing neutral atom traps, experiments cannot rely on slow spontaneous emission to obtain ground state antihydrogen \cite{kolb2021}. It is thus of paramount importance to either initially form strongly bound \Hbar quantum states (by for example acting, in the case of a three-body-reaction, on the \pbar-\pos plasma parameters as discussed in \cite{hunter2022}) or enhance the decay of nascent Rydberg \Hbar states in current experimental conditions. In view of the latter approach, several deexcitation schemes relying on either electric and magnetic field \cite{COM2019,COM2019_1} or light mixing of Rydberg states \cite{WOL20} associated with, in most cases, laser stimulated deexcitation have been theoretically identified. The techniques allow in principle to achieve close to unity ground state fractions of initially populated levels with principal quantum numbers $n\sim30$ within a few tens of microseconds.

We discuss here the concept and status of a hydrogen proof-of-principle experiment to commission stimulated deexcitation  techniques for application in antihydrogen experiments.

\section{Hydrogen proof-of-principle beamline}
\label{sec: H_proof_of_principle}

Due to the scarcity of antihydrogen atoms we have developed and built a hydrogen beamline to test and commission deexcitation techniques for application in experiments located at the Antiproton Decelerator. The experimental setup and different avenues toward the production of an excited Rydberg beam are discussed in the following.

\subsection{Atomic hydrogen beamline}

\begin{figure}
\centering
\includegraphics[width=\columnwidth]{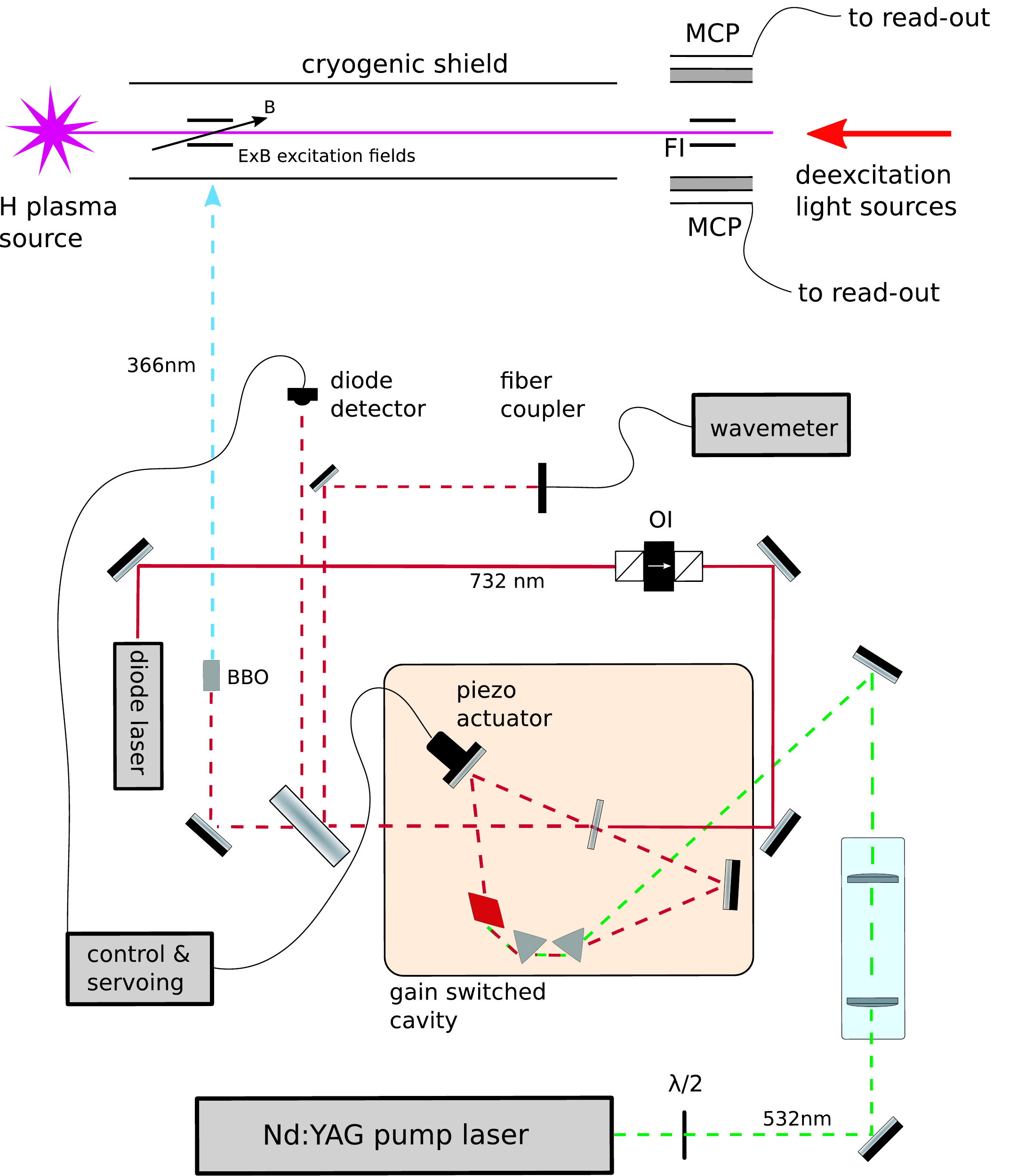}
\caption{Top: Schematic of the hydrogen beamline including the electric field ionizer (FI) and the microchannel plate (MCP) particle detectors. The electric and magnetic fields required for optical Rydberg excitation (cf.~section~\ref{subsubsec: coll-opt excitation}) yet need to be experimentally implemented. Bottom: Schematic of the injection seeded Ti:Sa Rydberg excitation laser.}
\label{fig: exp_scheme}   
\end{figure}

\begin{figure}
\centering
\includegraphics[width=\columnwidth]{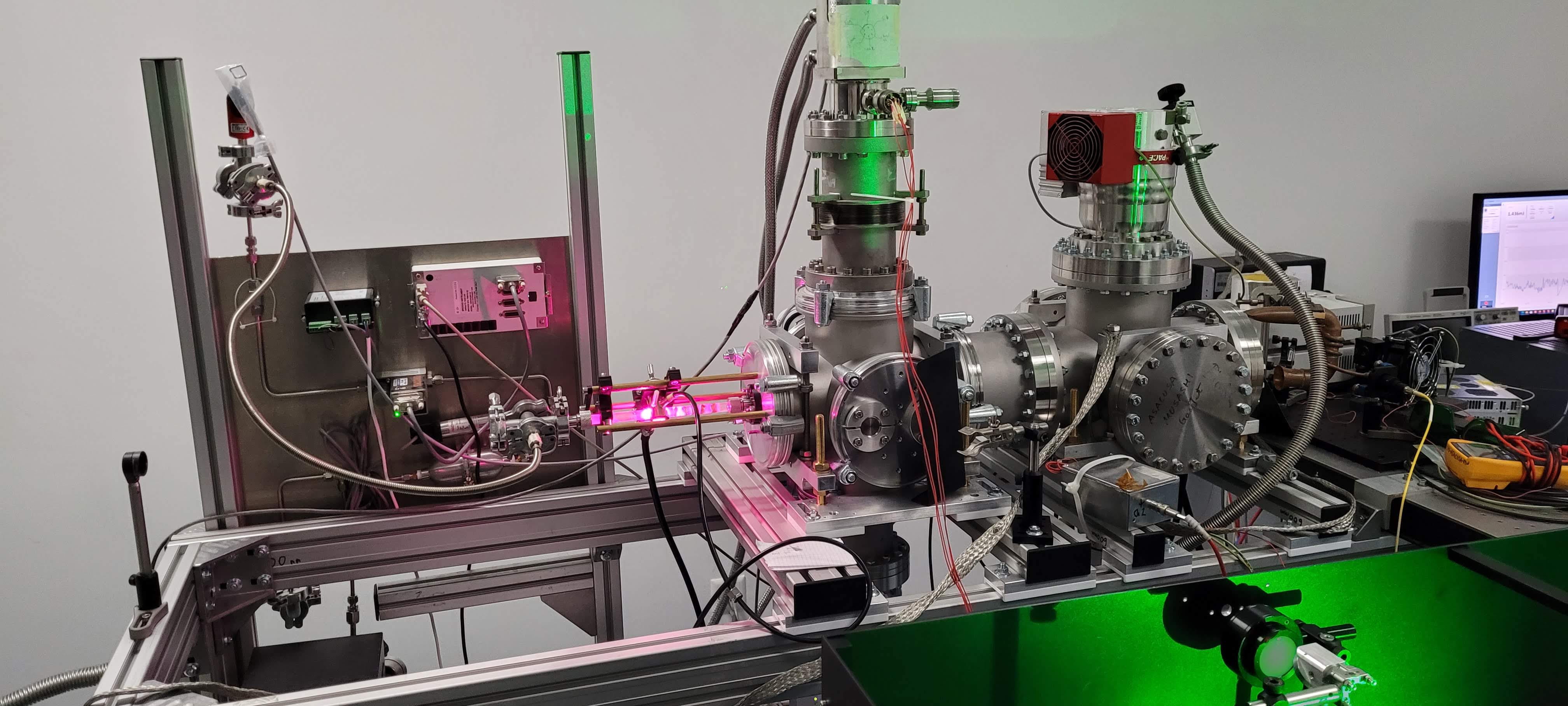}
\caption{Photograph of the hydrogen beamline. The molecular gas supply and the pink plasma discharge sustained by the Evenson cavity are shown on the left. Further downstream the beam, two vacuum chambers host the cryogenic beam shield and the instrumentation required to probe the beam's quantum state distribution, respectively.}
\label{fig: beamline_picture}
\end{figure}

The setup consists of a discharge plasma source to produce a thermal beam of atomic hydrogen \cite{malbr2019}. For this purpose, ultra-pure molecular hydrogen gas is produced from de-ionized water with an electrolysis generator. The gas is guided through a supply line consisting of a small buffer reservoir and a flow-controller from where it reaches, further downstream, a cylindrical quartz tube that is encased by a metallic resonator cavity. The latter (design \#5 in \cite{Evenson1965}) can be hand-tuned to efficiently sustain a microwave discharge plasma at some tens of watts allowing to dissociate the flowing molecular gas. Hydrogen atoms are emitted through a small pinhole into the downstream vacuum region. A cryogenic beam shield that is cooled with compressed helium to temperatures around \SI{25}{\kelvin} prevents thermal radiation from impacting the quantum state distribution in the atomic beam which can be probed further down the atoms' flight path. Upon exiting from the thermal shield, the beam enters an electric field region generated by two ionization meshes that are mounted parallel to the beam propagation direction at a distance of \SI{5}{\milli \meter}. The ionization products are accelerated toward and collected into the nearby MCP chevron stacks.
The quantum state distribution of the beam is investigated by counting the ionization events per time interval as a function of a varied electric field ionization strength. Rydberg state mixing and deexcitation light sources can illuminate the atomic sample through a dedicated vacuum window at the rear-end of the setup. An illustration of the beamline is provided in the top part of Fig.~\ref{fig: exp_scheme}. A photograph of the installation is shown in Fig.~\ref{fig: beamline_picture}.

\subsection{Hydrogen Rydberg beam production}

The source produces in large majority atoms in ground-state. In order to develop deexcitation techniques it is thus required to excite a substantial fraction of the atomic beam toward a Rydberg state. We follow different avenues first of which is a collisional-optical excitation as discussed in subsection \ref{subsubsec: coll-opt excitation}. Alternatively, Rydberg levels can be populated via recombination processes and electron impact inside the discharge plasma. We present, in section \ref{subsubsec: Rydberg from plasma}, results of a scan of the quantum state distribution of the beam emitted from the hydrogen source and discuss the complementarity of both approaches for our purpose.

\subsubsection{Collisional-optical excitation}
\label{subsubsec: coll-opt excitation}

\begin{figure}
\centering
\includegraphics[width=0.75\columnwidth]{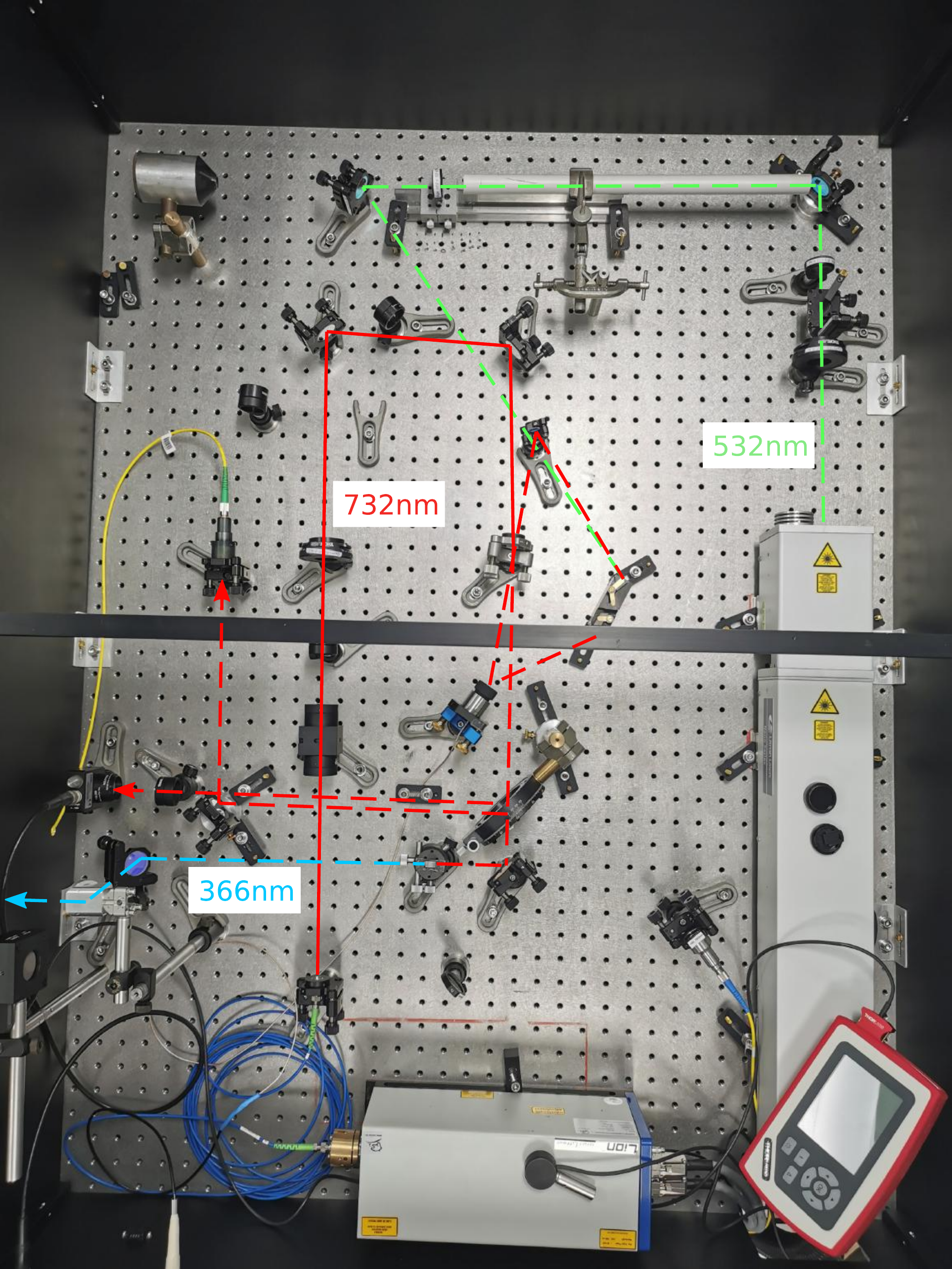}
\caption{Photograph of the Rydberg excitation laser setup. The laser beam paths are shown in analogy to the illustration shown in Fig.~\ref{fig: exp_scheme}. The picture is reprinted from \cite{WOL2022}.}
\label{fig: laser_picture}   
\end{figure}

We have developed and commissioned a laser capable of exciting 2s metastable atoms (radiative lifetime of \SI{0.12}{\second}) to highly excited Rydberg states with $n\sim30$. The setup is inspired by the work presented in \cite{drag2019}. A commercial frequency doubled Nd:YAG laser provides light pulses with a width of \SI{10}{\nano \second} and maximum average pulse energies of \SI{200}{\milli \joule} at \SI{532}{\nano \meter}. The pump beam is guided onto a Ti:Sa crystal inside a gain switched cavity. A fraction of the resulting red laser light impinges on a detection diode and a wavemeter while the main beam is again frequency doubled within a BBO crystal to obtain the required 2s$\, \rightarrow \,$30p transition wavelength of \SI{366}{\nano \meter}. In order to achieve a narrow spectral emission profile, the cavity is injection seeded with a few \si{\milli \watt} cw laser diode at the desired wavelength. The cavity is kept at resonance with the seeding beam relying on a piezo-electric mirror mount to compensate for drifts and therefore establish stable single-mode emission. The piezo-element is controlled with an electronic box and associated software that is based on the work presented in \cite{PIDbox}. The installation is illustrated in the bottom part of Fig.~\ref{fig: exp_scheme} and a picture is shown in Fig.~\ref{fig: laser_picture}. The parameters of the optical setup and its light output characteristics at the mentioned wavelengths are specified in Table \ref{tab: laser_characteristics}. In the unseeded case, the light emission occurs in both directions along the resonator cavity, whereas the pulse energy in single-mode operation is concentrated into the forward direction only. This results in an approximate doubling of the output average pulse energy when the cavity is injection seeded. The maximum achievable \SI{366}{\nano \meter} average pulse energy is found to be \SI{10.1\pm0.9}{\milli \joule}. This parameter is limited by the damage threshold of the Ti:Sa crystal which is reached at $\sim \,$\SI{30}{\milli \joule} average pulse energy in the \SI{732}{\nano \meter} beam just downstream the cavity.

\begin{table}
\centering

\begin{tabular}{|c|c|c|c}
\hline
\multicolumn{2}{|c|}{pulse repetition rate} & \multicolumn{2}{|c|}{\SI{10}{\hertz}} \\
\multicolumn{2}{|c|}{pulse duration} & \multicolumn{2}{|c|}{\SI{10}{\nano \second}} \\
\multicolumn{2}{|c|}{cavity length} & \multicolumn{2}{|c|}{\SI{0.65}{\meter}} \\
\multicolumn{2}{|c|}{output coupler reflectivity} & \multicolumn{2}{|c|}{\SI{20}{\percent}} \\
\multicolumn{2}{|c|}{Ti:Sa crystal dimensions} & \multicolumn{2}{|c|}{5~x~10~x~20~\si{\cubic \milli \meter}} \\
\multicolumn{2}{|c|}{BBO crystal dimensions} & \multicolumn{2}{|c|}{5~x~5~x~10~\si{\cubic \milli \meter}} \\
\multicolumn{2}{|c|}{beam waist (532/732/366~nm)} & \multicolumn{2}{|c|}{3.5/2/2~mm} \\
\multicolumn{2}{|c|}{injection diode cw power} & \multicolumn{2}{|c|}{\SI{5}{\milli \watt}} \\
\multicolumn{2}{|c|}{injection diode beam diameter} & \multicolumn{2}{|c|}{\SI{2}{\milli \meter}} \\
\multicolumn{2}{|c|}{single-mode locking efficiency} & \multicolumn{2}{|c|}{\SI{97}{\percent}} \\
\hline
\end{tabular}

\bigskip

\begin{tabular}{|c|c|c|c|}
\hline
\multicolumn{4}{|c|}{Average laser pulse energy [mJ]} \\
\hline
532~nm & $107.2\pm1.3$ & $127.3\pm1.3$ & $145.1\pm1.3$ \\
\hline
732~nm & $\mathbf{14.7\pm1.4}$ & $\mathbf{22.5\pm2.4}$ & $\mathbf{29.6\pm3.9}$ \\
\hline
366~nm  & $1.7\pm0.2$ & $2.9\pm0.2$ & $4.1\pm0.3$ \\ 
        & $\mathbf{3.9\pm0.6}$ & $\mathbf{7.1\pm0.8}$ & $\mathbf{10.1\pm0.9}$ \\
\hline
\end{tabular}

\caption{Top: Summary of the parameters of the optical installation. Bottom: Average laser pulse energies for the pump beam at \SI{532}{\nano \meter}, the Ti:Sa cavity output at \SI{732}{\nano \meter} and the frequency doubled BBO laser output at \SI{366}{\nano \meter}. The values are displayed in bold font for the injection seeded and in regular font for the unseeded case, respectively.}
\label{tab: laser_characteristics}
\end{table}

The techniques proposed for stimulated deexcitation allow to address the entire distribution of quantum states formed in a typical antihydrogen experiment -- most importantly including the preferentially populated and longest-lived high $(n,l,m)$ levels with lifetimes $\tau \propto n^3l^2$. We are thus aiming at producing a beam of Rydberg atoms containing these states of interest. To this goal, we are in the process of designing coils and electrodes to allow for optical excitation, using the laser discussed above, toward circular states via the crossed fields method discussed and experimentally demonstrated in \cite{morgan2018,Lutwak1997}.

The optical transition stimulated by the laser discussed above requires a metastable 2s beam which can be generated from ground-state relying on different approaches. Optical excitation toward the 3p levels allows to obtain 2s states according to the spontaneous decay branching ratio of \SI{12}{\percent} \cite{Harvey1982}. Another possibility lies in a two-photon 1s$\, \rightarrow \,$2s excitation or electron collisional processes \cite{Biraben1990, Stebbings1960}. For the latter technique we have developed and currently commission an electron gun. The beamline is designed such that the resulting hydrogen beam deflections caused by electron impact can be compensated for.

Following the observation of Rydberg states produced through collisional and recombination processes inside the electron discharge plasma, as discussed in the following part, metastable 2s atoms can be expected to emerge either from the source through electron collisions inside the plasma or population through spontaneous decay from higher excited states downstream the plasma inside the beam. We could, as of now, see no clear evidence of the 2s state being populated in the beam when detecting Lyman-alpha photons with the MCPs and relying on electric field quenching. Possible explanations involve the depopulation via 2p states already close to the plasma within the microwave cavity (indeed the few tens of \si{\watt} required to sustain the discharge on a few \si{\centi \meter} result in electric fields of some ten \si{\volt \per \centi \meter}) or just upstream the field-ionization region caused by stray fields of the MCPs which would result in a reduced detection efficiency.

\subsubsection{Emission of excited states from the electron discharge plasma}
\label{subsubsec: Rydberg from plasma}

We performed simulations relying on the collisional radiative model Yacora \cite{Wund2020} to theoretically investigate the production, in the microwave discharge, of highly excited Rydberg states that are of interest for studies of stimulated deexcitation. We estimate typical electron densities in our setup to be $10^{14} \, \si{\per \cubic \meter} \leq n_e \leq 10^{15} \, \si{\per \cubic \meter}$ \cite{chabert2011}. Electron temperatures $T_e$ can be determined by measuring a spectral line emission ratio (eg. H-$\alpha$/H-$\beta$). A comparison with theory then leads to $T_e<1 \, \si{eV}$ \cite{kulk2021}.

There exist many recombination and excitation processes from the ground state that can lead to the population of quantum states in the vicinity of $n\sim30$. The list of reactions implemented in Yacora can be found in \cite{yacora_online} or the above cited reference. We find that mainly recombination processes of ionic species determine the population density of highly excited Rydberg manifolds. The population of stronger bound levels, in contrast, gets quite rapidly dictated by collisional excitation from the ground state. This is a typical observation in such a so-called recombining plasma regime where $T_e < 1 \, \si{eV}$ \cite{ThesisGiacomin}. Extracting the exact contribution of each simulated channel to the Rydberg population coefficients remains however difficult due to the lack of knowledge of the involved ion densities and temperatures. Further diagnostic tools, like a Langmuir probe to precisely assess the electron temperature and most importantly the density of different ionic species in the discharge, would be needed to determine the relative importance of the different simulated excitation and recombination processes in the plasma.

\begin{figure}
\centering
\includegraphics[width=\columnwidth]{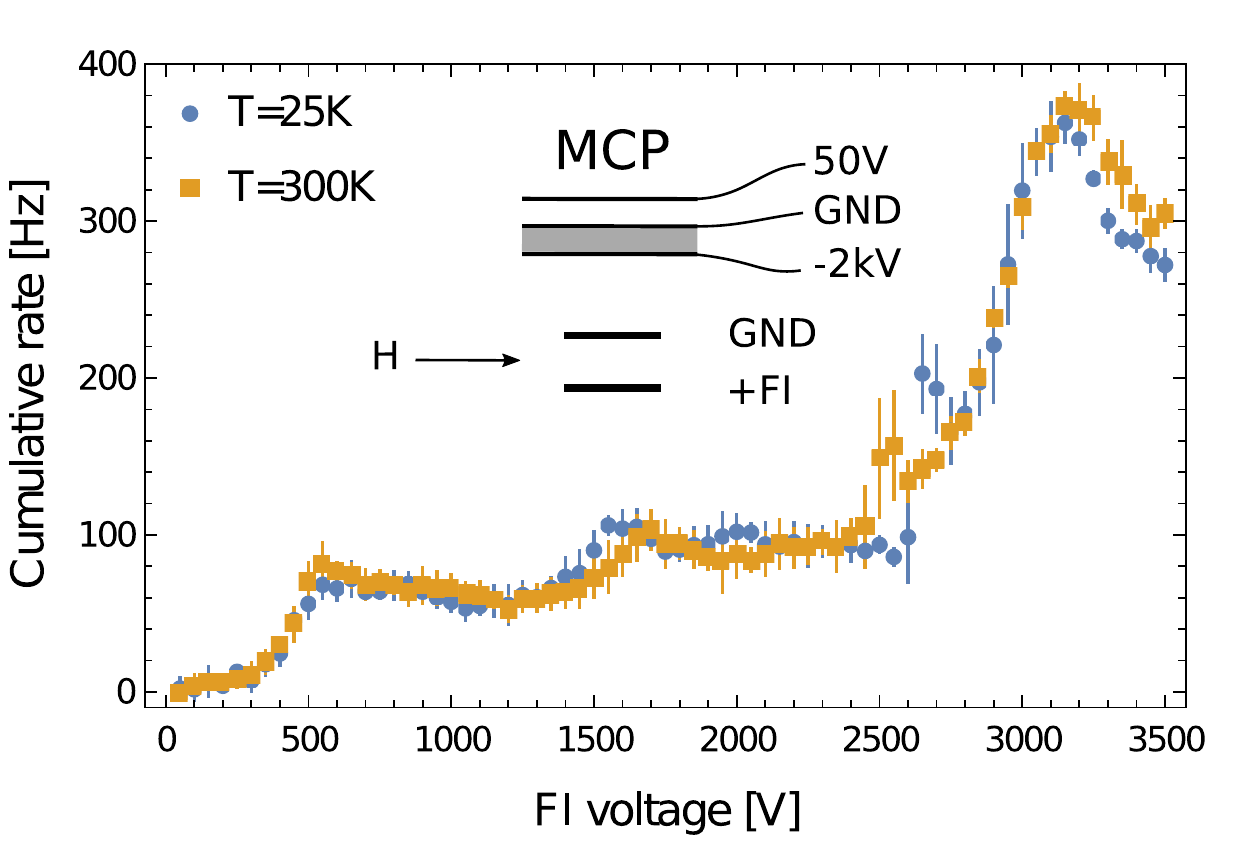}
\includegraphics[width=\columnwidth]{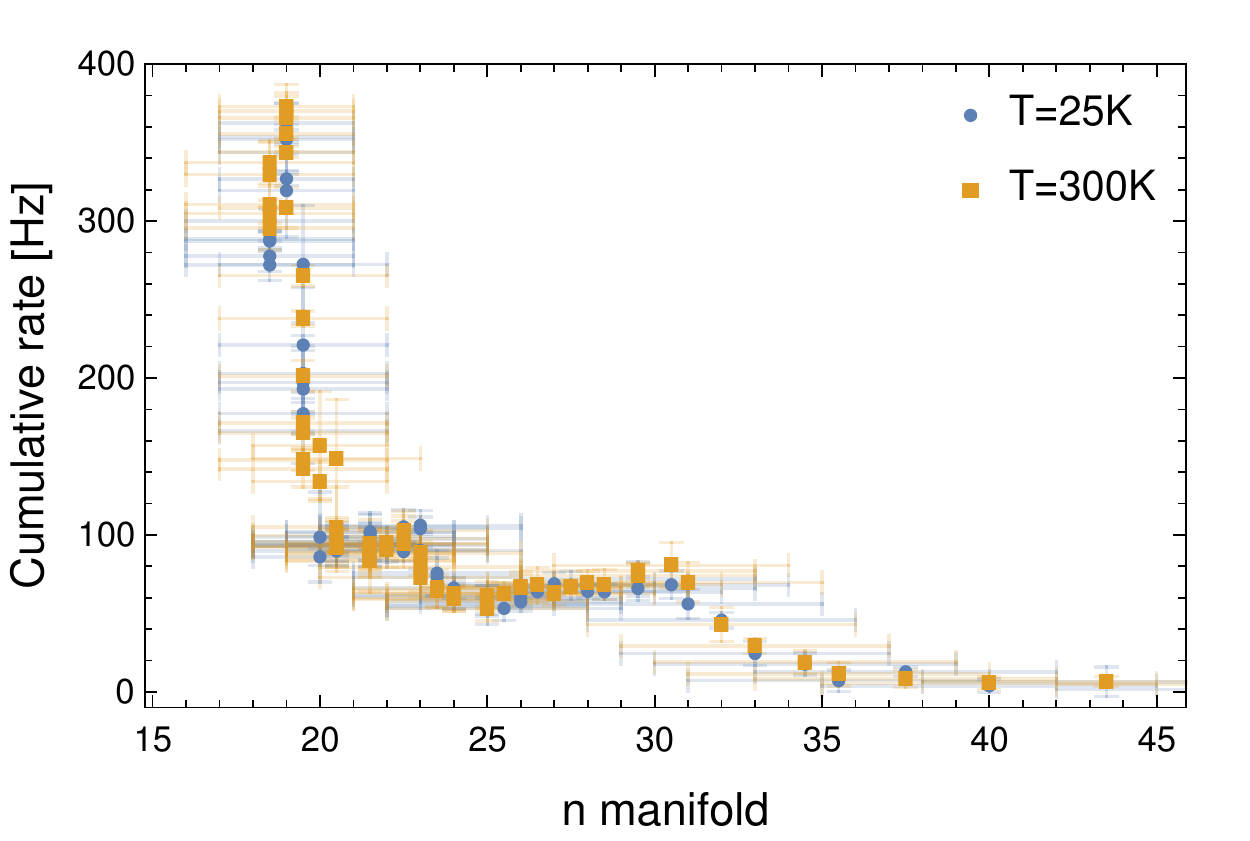}
\caption{Top: Background subtracted rate of detected protons originating from the ionization of Rydberg hydrogen atoms as a function of the voltage applied to the field ionization meshes. Bottom: The same signal plotted as a function of the range of the atoms' principal quantum numbers $n$ probed at each voltage setting (horizontal bars reflect the lack of knowledge of the distribution of substates within each $n$-manifold). We take into account a $\pm \,$\SI{0.5}{\volt} error on the potential applied to the field ionizer meshes (mounted \SI{5}{\milli \meter} apart from each other) and a \SI{10}{\percent} imprecision in the spacing between the potential grids. Both plots show results for two different temperatures of the cryogenic beam shield.}
\label{fig: FI_scans}   
\end{figure}

We measured the distribution of the hydrogen quantum states emitted from the microwave discharge and detected a few hundred \si{\hertz} increase in the count-rate of protons emerging from the atomic ionization process as a function of the electric field strength established between the ionizer meshes. The protons were detected in a single MCP configuration. The detector output was amplified and digitized. The events were discriminated as a function of the peak voltage of the few ten nanoseconds long charge burst detected on the anode of the MCP. The results and the potentials that were applied to the ionization meshes and the MCP are shown in Fig.~\ref{fig: FI_scans} (top).

Each field ionizer setting probes a range of $n$-manifolds (for more details and formulas, please refer to chapter 5 of \cite{kolb2021}). A graph of the proton detection rate as a function of the corresponding range of $n$ (horizontal bars) is shown in Fig.~\ref{fig: FI_scans} (bottom). Even though the data cannot yet be compared in a quantitative way to simulation results, qualitative features of the Rydberg state distribution can be extracted. The sudden drop for $n>30$ can be explained by ionization most likely already within the few \SI{10}{\volt \per \centi \meter} stray field close to the microwave discharge region. The presence of states with $20 \leq n \leq 30$ is of high interest for deexcitation studies, especially because we expect the population of a large number of $n$-manifolds, which would reproduce conditions similar to those found in antihydrogen experiments. The scans are sensibly the same at room temperature and \SI{25}{\kelvin}, where the blackbody irradiance in the frequency regime critical for ionization is reduced by more than an order of magnitude compared to \SI{300}{\kelvin}. We conclude that the quantum states present in the beam are not very susceptible to being coupled to the continuum via THz radiation which is characteristic for high angular momentum states that exhibit minimum ionization cross sections (cf.~Fig.~11 in \cite{WOL20}). In fact, one can indeed expect that non-circular states emerging from the plasma rather rapidly collapse to circular levels along their spontaneous decay cascade \cite{Flannery2003}.
The lifetime $\tau$ of a $(n,l)$ state with magnetic quantum number $|m| \leq l < n$ can be approximated by \cite{chang1985}
\begin{equation}\tau = \left( \frac{n}{30} \right)^3 \left( \frac{l+1/2}{30} \right)^2 \times 2.4 \, \si{\milli \second}.\label{eq:lifetime}\end{equation}
For beam velocities of the order of some \SI{1000}{\meter \per \second} and the \SI{0.7}{\meter} long flight path from the discharge plasma to the detection region, quantum states with lifetimes of the order of \SI{100}{\micro \second} can be expected to spontaneously decay before reaching the field ionization region (for $(n,l)=(16,15)$ one finds $\tau\sim97\,\si{\micro \second}$, cf.~Eq.~\ref{eq:lifetime}). Consequently, the signal levels off toward the low lying $n$ states. At the same time, electron collisional processes start to play an increasingly important role toward stronger bound states which explains the intermittent rapid increase in rate between $15<n<20$.

For settings above $\sim \,$\SI{3.2}{\kilo \volt}, the cumulative detection rate drops from close to \SI{400}{\hertz} back to roughly \SI{300}{\hertz} at \SI{3.5}{\kilo \volt}. Part of an explanation for this might be the ionization of weakly bound atoms upstream the detection region by the increasingly large stray field emerging from the ionization stage. These protons must be expected to be less efficiently detected which can lead to a net drop in count-rate on the MCP at high ionization field strengths under the assumption that no additional strongly bound states are accessed through the voltage ramp-up. Consequently, the region around and beyond $\sim \,$\SI{3.2}{\kilo \volt} constitutes the maximal field ionization strength that can be employed and data at such high voltage configurations must be handled with care since this process might build up gradually.

\section{Summary and outlook}

We reported on the design and on the status of a proof-of-principle experiment to demonstrate the performance of deexcitation techniques for antihydrogen atoms. An atomic hydrogen beam and an excitation laser have been developed to excite metastable 2s atoms toward Rydberg levels. An electron gun for the production of 2s states from the ground-state atoms emitted out of the microwave discharge plasma is currently being commissioned. We are in the process of designing a circular Rydberg production stage to introduce the required crossed electric and magnetic fields into the laser excitation region. This will allow for the production of single quantum states of interest for deexcitation studies.

In parallel, the production of a broad distribution of highly excited atoms with $20 \leq n \leq 30$ via recombination and collisional processes inside the discharge plasma has been experimentally evidenced relying on electric field ionization. This approach best reproduces the conditions faced in an antimatter experiment. We point out that this result can be of interest to the plasma community to, for example, benchmark collisional-radiative models in the high $n$-regime.

Suitable (anti)hydrogen mixing and deexcitation light sources have been tested in a cesium proof-of-principle experiment in \cite{vieil2020} and in particular photomixing has been identified as a versatile and promising technology for our deexcitation purposes. The forthcoming step is the demonstration of fast stimulated Rydberg state mixing and deexcitation in hydrogen. We plan on developing a tailored photomixing device for the application in (anti)hydrogen experiments in the coming months.

The reported developments lay the foundation for a first stimulated deexcitation result in the near future. We then aim, in a subsequent step, for a swift installation of the commissioned and optimized technology in experiments at CERN's Antiproton Decelerator to enable the production of ground-state antihydrogen required to perform gravity and spectroscopy measurements in beam configurations.

\section*{Acknowledgements}
This work has been sponsored by the Wolfgang Gentner Programme of the Bundesministerium für Bildung und Forschung (Grant No. 05E15CHA, university supervision by Norbert Pietralla) and the Austrian Science Fund (FWF): W1252-N27. It was supported by the Studienstiftung des Deutschen Volkes. In addition, the research presented was financed by the LASANTI grant (from Domaine d'Int\'{e}r\^{e}t Majeur Science et Ing\'{e}nierie en R\'{e}gion \^{I}le-de-France pour les Technologies Quantiques -- DIM SIRTEQ --) and the  Initiative de Recherche strat\'{e}gique of the Universit\'{e} Paris-Saclay IQUPS (project number 156541). We warmly thank P. Chabert and C. Drag for useful discussions concerning radio-frequency plasma sources and the EP-AGS, EP-DI, EP-SAM, EN-AA and EP-ESE groups at CERN for the support provided for the laser room construction.

\bibliography{bib}
\end{document}